\documentstyle[epsfig]{article}
\begin{document}
\title{One Mass-Radius relation for Planets, White Dwarfs and Neutron Stars}
\author{D. Lynden-Bell\footnote{Institute of Astronomy, The Observatories, Madingley Road,
Cambridge, CB3 0HA}\ \& J.P. O'Dwyer\footnote{Dept. of Physics, Science Laboratories, University of Durham, Durham, DH1 3LE}}
\maketitle 
\begin{abstract}
We produce the simple formula for the radius, $R$, of a cold body:
$$\left ({\scriptstyle{4\over 3}}\pi\rho_o\right )^{1/3} R= M^{1/3}_p
{y  \over 1+ y^2 }I $$ where $y^3=M/M_p$, $M$ is the total mass, $J =
\sqrt{1-y^4/y^{4}_{ch}}$ and $I = J/ \left [1+{3\over 4 }\left
(1-J^{\scriptstyle{1\over 2}}\right )\right ]$.

$y^3_{ch}=M_{ch}/M_p$ where $M_{ch}$ is Chandrasekhar's limiting mass.
The density of our liquid/solid matter at low pressure defines the
constant $\rho_o$.  For bodies with non relativistic electrons
$I=J=1$.

$M_p$ is constant and gives the mass of the planet of maximum
radius.  $M_{ch}$, $M_p$ and $\rho_o$ are all defined in terms of
fundamental constants, in particular the ratio $M_p/M_{ch}$ is
basically given by the three halves power of the fine structure
constant.

The aim of the discussion is to emphasise physical principles and to
connect the masses derived to the fundamental constants, so
mathematical niceties are sacrificed and replaced by interpolations
between simple exact limits.
\end{abstract}
\section{Introduction}
The virial theorem shows that a star deprived of any energy sources
will shrink and its kinetic energy will grow which normally implies it
will get hotter.  In the 1920's Eddington (1926) was puzzled as to how a star
could ever cool down and end its life.  This problem, together with
the mystery of white dwarfs such as Sirius B, was beautifully solved
by R. H. Fowler (1926) who realised that the electrons would become
degenerate, as in a metal, so the star could then be supported by the
degeneracy pressure which is an inevitable consequence of the
zero-point energy combined with the Pauli exclusion principle.

Even bodies of zero temperature have such a pressure which supports
the white dwarfs against gravity.  Stoner (1929) made early models of white
dwarfs to compare their densities with the theory but Anderson (1929) rightly
criticised him for not allowing for the relativistic motions of
electrons forced to very high densities.  He showed that when
this was allowed for, there would be a limiting mass beyond which no
cold white dwarf models could exist.   His estimate of the limiting
mass was not very good, but Stoner (1930) then modified his own calculations
and calculated the limit for homogeneous models which gave an answer
good to 10\%.  Meanwhile, Chandrasekhar (1931) had developed the theory
independently starting from Fowler's work.  He realised that the
limiting configuration would be an $n=3$ polytrope and gave the first
accurate determination of the Chandrasekhar limit.

More detailed models of white dwarfs were given by Salpeter (1967) who
derived conditions for their solidification and determined properties
of the lattice.  Shapiro and Teukolsky in their book give a definitive
account of how all this extends to neutron stars.

Here our aim is much more modest.  To give a rough approximate formula
for the radius of any cold body as a function of its mass.  We
illustrate how the balance between degeneracy pressure, the electrical
attraction of electrons to atomic nuclei, and gravitation lead to a
mathematical formula that describes the mass-radius relationship for
cold bodies.  This holds from asteroids, through planets and brown
dwarfs to white dwarfs and neutron stars.  Understanding rather than
accuracy is our aim so we shall use interpolation formulae where they
are simpler, nevertheless we shall ensure accuracy in important
limiting cases.  Thus we aspire to follow the fine tradition set in
Salpeter's 1967 lectures.  

At masses below that of Saturn, the degeneracy pressure of the
electrons within  the atoms (vide the Thomas Fermi atom) is balanced
by the electrical binding.  While the gravity serves to hold the whole
body in a nearly spherical shpae, it is not strong enough to crush the
atoms into significantly smaller volumes.  Thus the material is nearly
incompressible and apart from any chemical fractionation the lighter bodies have the same density.  However, above the
mass of Saturn the gravitational potential energy increases to become
comparable with the electrical one.  Above the mass of Jupiter, the
gravitation dominates, the atoms are crushed together with the outer
electrons eventually bcoming pressure ionised to make a continuous
sea.  In fact, somewhat above Jupiter's mass, the weight of each layer
added so compresses the material within that the whole body shrinks in
radius.  The electron sea is not so dense that the zero point energy
drives the electrons to speeds close to that of light, so they give
Fowler's $p=K\rho^{5/3}$ as an equation of state as in $n=3/2$
polytropes. The $\int pdV$ energy behaves as $M^{5/3}/R^2$.  Although
the $-GM^2/R$ gravitational energy rises with a higher power of the
mass, the total energy achieves its minimum at the (equilibrium) radius
with $R\propto M^{-4/3}$.  This is the brown dwarf white dwarf
sequence.  However, as $M$ is further increased the increasing density
of the bodies $(\bar{\rho}\propto M^2)$ leads to zero point agitation
with relativistic speeds.  This results in the pressure density
relationship $p=K^\prime\rho^{4/3}$ with  $\int pdV \propto M^{4/3}/R$.  As
this is approached the equilibrium radius rapidly becomes very small
and very sensitive to mass because both gravitational and internal
energies scale with the same power of the radius.  No force balance is
possible beyond Chandrasekhar's limiting mass.

In neutron stars the electrical energy is unimportant; most of the
electrons have combined with protons and been transformed into
neutrons so that neutron degeneracy supplies the pressure.  Neutron
star radii are smaller than white dwarf radii by the factor ${m_e
\over m_n}$ so the densities are $(m_n/m_e)^3$ times greater.

A secondary aim of this article is to express the important masses
that arise in astronomy in terms of the fundamental physical
constants.  Thus the ratio of the Chandrasekhar mass to the mass of
the planet of maximum radiu is essentially the fine structure constant
raised to the power $-3/2$.  Such matters are of particular importance
to the misguided few who seriously entertain theories in which such
fundamental constants change.

\section{Kinetic Energy of Degenerate Material}
The kinetic energy of an electron is given by 
$$\epsilon = m_e c^2 \left [\sqrt{ {p^2\over m^2_e c^2} +1 } -1 \right
]\ \ . \eqno (1)$$
The number of energy levels in a box of side $L$ with momentum between
$p$ and $p+dp$ is 
$$2 \times 4\pi p^2dp L^3 h^{-3}\ \ , $$
where the initial 2 comes from the 2 spin states of an electron.  If
all levels up to momentum $p_m$ are filled, the kinetic energy of the
electrons per unit volume is
$${\cal E} = \int^{p_m}_o 8\pi h^{-3}{\epsilon} p^2 dp \ \ . \eqno (2)$$
While the integral may be evaluated using $sh^{-1}\left ({p_m \over
m_ec}\right )$ that begins to obscure the physics so we shall evaluate
the limiting cases and interpolate.  

If $p_m \ll m_ec$ then ${\epsilon} = {1 \over 2}p^2/m_e$ so
$${\cal E} = {{\scriptstyle 4}\pi \over {\scriptstyle 5}}
h^{-3}p^5_m/m_e \ \ . \eqno (3)$$ If $p_m \gg m_ec$ then ${\epsilon}
\simeq pc$ for most electrons, so
$${\cal E} = 2\pi h^{-3}p_m^4 c \ \ . \eqno (4)$$ Looking at (1), an
interpolation in terms of the function $\sqrt{x^2+1} -1$ where $x=\eta
{p_m \over m_ec}$ suggests itself, where $\eta$ is to be determined.
For large $x$ this function gives $\eta p_m/(m_ec)$ while for small
$x$ it gives ${\scriptstyle {1\over 2}}\eta^2 {p^2_m \over m^2_ec^2}$.

So the ratio of the non-relativistic case to the very-relativistic
case is ${\scriptstyle{1 \over 2}} \eta {p_m \over m_ec}$.  Now the
ratio of the expressions (3) and (4) is ${\scriptstyle{2\over 5}}\
{p_m \over m_ec}$ so we choose $\eta = {\scriptstyle{4\over 5}}$.  The
resulting interpolation formula exact at both limits is
$${\cal E} = {\scriptstyle {5 \over 2}} \pi h^{-3}p^3_m m_ec^2 \left
(\sqrt {x^2 +1} -1 \right ) \ \ , \eqno (5)$$
where
$$x= {\scriptstyle{4\over 5}} {p_m \over m_ec}\ \ . \eqno (6)$$ 
If the total mass per electron is $\mu_e m_H$ then the density is
$$\rho = \mu_em_H {\scriptstyle{8\over 3}}\pi p^3_m h^{-3}=\rho_1
x^3\ \ , \eqno (7)$$ where
$$\rho_1 = \left ({\scriptstyle{5 \over 2}} \right )^3 {\pi\mu_em_H \over
3}/\left [h/(m_ec)\right ]^3 \ \ ,\eqno (8)$$
thus $\rho_1$ is the density at which the electrons start to become
relativistic due to the uncertainty principle.  $m_H$ is the mass per
electron and $h/(m_ec)$ is its Compton wavelength.  Rewriting (5) in
terms of $\rho$
$${\cal E}/\rho = {\scriptstyle{15 \over 16}} {m_ec^2 \over
m_H\mu_e}\left (\sqrt {x^2+1} -1 \right ) \ \ . \eqno (9)$$ Now
imagine expanding unit mass of this material
$$P dV=Pd\left ({{\scriptstyle 1}\over \rho}\right )=-d\left ({\cal
E}/\rho \right ) = -d\left ({\cal E}/\rho \right )/dx \ dx \ \ . \eqno
(10)$$ Since $\rho \propto x^3$ we find the pressure density
relationship
$$P=P_1 {x^{\scriptstyle 5} \over {\sqrt{x^2+1}}} \ \ ,\eqno (11)$$
where
$$P_1={{\scriptstyle 5}^4 \over {\scriptstyle
2}^7}{\pi \over 3} {m^4_e c^5 \over h^3} \ \ . \eqno (12)$$ 
In particular the non-relativistic and extreme relativistic limits
give $P=K_\Gamma\rho^\Gamma $,

\bigskip
\begin{tabular}{llllr}
$\begin{array}{l}
\\
P = K_\Gamma \rho^\Gamma = \\
\\ \end{array}$
&
	$\left\{\begin{array}{l}   
	K_{5/3} \rho^{5/3} \\
	\\
	K_{4/3} \rho^{4/3} \end{array}\right.$    
&			
		$\begin{array}{l}\Gamma = 5/3 \\
		\\
		\Gamma = 4/3\end{array} $
&
			$\begin{array}{l} 
			{\rm non-rel} \\
			\\
			{\rm extreme\ rel.}\end{array}$	
&
				$\begin{array}{r}
				\\
				(13)\\
				\\
				\end{array}$ \\
& & & & \\
\multicolumn{3}{l}{From (11) and (7),} & \\
& & & & \\
$\begin{array}{l}
\\
K_\Gamma = \\
\\ \end{array}$
&
	$\left\{\begin{array}{l}   
	P_1 \rho^{-5/3}_1 \\
	\\
	P_1 \rho^{-4/3}_1 \end{array}\right.$    
&
$\begin{array}{l}\Gamma = 5/3 \\
		\\
		\Gamma = 4/3\end{array} $
&
			$\begin{array}{l} 
			\\
			\\
			\end{array}$	
&
				$\begin{array}{r}
				\\
				\hspace{0.8cm}(14)\\
				\\
				\end{array}$\\
& & & & \\
\multicolumn{3}{l}{notice $P_1 /\rho_1 = {5 \over 16}\ \ \ {m_e
\over \mu_em_H}c^2 \ \ .$} & \\
& & & & \\
\end{tabular}

Returning to (10) we see that whenever ${\cal E} \propto \rho^\Gamma$ then 
$$P = (\Gamma -1){\cal E} \ \ . \eqno (15)$$

We have of course devised our derivation so that these formulae (8),
(13), (14) \& (15) are exact $cf$ Shapiro \& Teukolsky (1983).

Three limiting density distributions are important

\noindent
1. The uniform density which we shall see occurs.

\noindent
2. The non-relativistic $\Gamma = {\scriptstyle{5\over 3}}$ polytrope of index
   ${\scriptstyle{3\over 2}}=n={1\over \Gamma -1}$\ \ .

\noindent
3. The extreme relativistic polytrope of index $3, \Gamma = {\scriptstyle{4\over 3}}$.

To orient ideas it is useful to consider initially the uniform density
case.  The kinetic energy is given by (9) \& (7) 
$$\mathrm{T} = {\scriptstyle{4 \over 3}}\pi R^3{\cal E} = A_oM\left [\sqrt{
(\zeta_oM^{1/3}/R)^2+1} -1 \right ] \ \ , \eqno(16)$$ 
where we have written $x = \zeta_oM^{1/3}/R$ and $A_o = {15 \over 16}
\ \ {m_e c^2 \over m_H\mu_e} = {3P_1 \over \rho_1}$, 
$$\zeta_o =\left ({\scriptstyle{4 \over 3}}\pi \rho_1 \right )^{-1/3}\ \
. \eqno (17)$$
Although this uniform case is only realistic for very small $x$,
nevertheless the modifications needed for polytropic cases occur only
via structure constants of order unity.  Thus much understanding can
be gained by studying the problems with $\mathrm{T}$ given by (16) and
(17) for all $x$.  To show the modifications needed for the polytropic
cases we now take the case of a polytrope of index $n = {1 \over
\Gamma -1}$.  Using (15) 
$$\mathrm{T}=\int^R_o4\pi r^2{\cal E}dr=nK_\Gamma R^3\int^1_o4\pi\left
({r \over R }\right )^2\rho^\Gamma d \left (r /R \right )=
nK_\Gamma M \left ({3M \over 4\pi R^3} \right )^{\Gamma -1}C_n \ \ ,
$$
where 
$$C_n = \int^1_o3\left ({r \over R }\right )^2 \left ({\rho \over
{\bar\rho}}\right )^{1+1/n}d\left (r /R \right ) \ \ , \eqno
(18)$$ is a structure constant for a polytrope of index $n$ which is
evaluated in the appendix: $C_o = 1, \ \ \ C_{3/2} = 1.7501, \ \ C_3 =
2.2146$.  Notice that our uniform density case (16) also gives the
$M(M/R^3)^{\Gamma -1}$ factor for both the non-relativistic $\Gamma =
{\scriptstyle{5\over 3}}$ and the extreme relativistic $\Gamma =
{\scriptstyle{4\over 3}}$ cases.  Following our earlier interpolation
procedure using $\eta $ below equation (4), we now look for an
interpolation suggested by (16) between the $\Gamma =
{\scriptstyle{5\over 3}}$ and $\Gamma = {\scriptstyle{4\over 3}} $
polytropes.  This is of the form
$$\mathrm{T}=AM\left[\sqrt{\left({\zeta M^{1/3}\over R} \right)^2+1}-1
\right] \ \ , \eqno (19)$$ 
where $A$ and $\zeta$ are chosen to make the two polytropic formulae
exact when $\zeta M^{1/3}/R \ll 1 \ {\rm and }\ \gg 1$ respectively.
This gives
$$\zeta = {C_{3/2}\over C_3}\zeta_o = 0.7903 \zeta_o;\ \ A\zeta^2 =
C_{3/2}A_o\zeta^2_o = 1.7501 A_o\zeta^2_o \ \ . \eqno(20)$$
Notice that (19) has precisely the same form as (16) but the values of
$A$ and $\zeta$ differ.

\section{Potential Energy    }
The gravitational potential energy is $-\alpha GM^2/R$ where Ritter's
formula tells us that for polytropes $\alpha = {3\over 5-n}$ (see
appendix).  However, for planets and small bodies the electrons are
held in electrically rather than gravitationally.  We must therefore
include in the potential energy some estimate of the electrical
potential energy.  The mean number density of electrons is
$$n_e=M/\left ({\scriptstyle{4 \over 3}}\pi R^3\mu_e m_H\right )\ \ ,$$
so we estimate the electrical potential energy as
$$-\beta \left({M\over \mu_e m_H}\right) \ {e^2\over R \left ({M\over
\mu_e m_H} \right )^{-1/3}} \ \ , \eqno (21)$$
where the first bracket gives the number of electrons in the whole
body and $\beta$ is a dimensionless constant to be determined.  The
gravitational potential energy is proportional to $M^2$ while the
electrical is proportional to $M^{4/3}$ at fixed $R$.  Particular
interest centres on the mass at which these two contributions are
equal.  Calling it $M_1$ we find with $\alpha = {\scriptstyle{3\over 5}}$
$$M_1 = \left (5 \beta /3 \right )^{3/2} \mu^{-2}_e \left ({e^2\over
Gm^2_H}\right )^{3/2}m_H = \left ({5\beta \over 3}\right
)^{3/2}\mu^{-2}_e 2.308 \times 10^{30}gm \ \ . \eqno (22)$$ Except for
a structure factor of order unity $M_1$ is given by $\left ({e^2 \over
Gm^2_p}\right )^{3/2}$, the ratio of the electrical to gravitational
forces between two protons, raised to the $3/2$ power, times $m_H$.
Notice that this mass does not depend on Planck's constant.  In
adapting formula (21) to rocky planets \& asteroids we have used
$\rho_o = 3.65$ and $\mu_e = 2$.

\section{Energy Minimisation Gives Radius}
The total energy is now given by 
$$E=AM\left (\sqrt{\zeta^2M^{2/3}R^{-2}+1}-1\right ) -\alpha {GM^2
\over R} \left [ 1 + \left ({M_1 \over M}\right )^{2/3}\right ]
\ \ .  \eqno (23)$$
The equilibrium radius has the minimum $E$ for any $R$ so 
$${R \over M} \ \ {dE \over dR} =0 \ \ , \eqno (24)$$
$${\rm therefore}\ {A\zeta^2M^{2/3}R^{-2} \over
\sqrt{\zeta^2M^{2/3}R^{-2}+1}} = {\alpha \over \zeta}\ \ {\zeta
M^{1/3} \over R} \ \ G\left (M^{2/3} + M_1^{2/3}\right ) \ \ ,
\eqno (25)$$
equation (25) is readily solved for $\zeta M^{1/3}R^{-1}$ and thence
for $R$
$$R={\zeta M^{1/3}\over q}\sqrt{1-q^2}\ \ , \eqno (26)$$
where
$$q={\alpha \over \zeta} \ {G\over A} \left (M^{2/3}+M_1^{2/3} \right
) \ \ . \eqno (27)$$ Evidently $R=0$ when $q=1$ so this gives
Chandrasekhar's limiting mass. Since $M_1^{2/3}$ is negligible at this
limit we have (putting $\alpha = {\scriptstyle {3 \over 2}}$ for the
$n=3$ polytrope $\Gamma ={\scriptstyle{4\over 3}}$)
$$M_{ch} = \left ({\zeta A \over \alpha G}\right )^{3/2} =
{\scriptstyle{3 \over 2}}{\sqrt \pi} \left ({C_3 \over 2}\right )^{3/2}
\left ({\hbar c \over Gm^2_H}\right )^{3/2} {m_H \over \mu^2_e} \ \
. \eqno (28)$$ Which demonstrates that the Chandrasekhar limit is
essentially the gravitational fine structure constant to the power of
$-{\scriptstyle{3\over 2}}$ times the mass of the hydrogen atom.
Equation (26) when combined with (27) gives the mass-radius
relationship
$$R={\zeta^2A \over \alpha G} \ \  {M^{1/3} \over M^{2/3} +
M^{2/3}_1} \sqrt{1-\left ({M \over M_{ch}}\right )^{4/3}}\ \ ,
\eqno (29)$$
where we have simplified the surd taking account of the fact the
$M_{ch} \gg M_1$.

Equation (29) shows that there is a maximum radius for a cold planet
close to $M_1$ i.e., close to Jupiter's mass.  Also, for small $M,\
R$ grows like $M^{1/3}$ so the density is constant.  This is not
surprising; the electrons are held in by electricity and we are merely
placing together electrically bound neutral objects.  As more mass is
piled on the growth in $R$ slows until close to $M_1$ the weight
of the overlying material so crushes what is beneath that additional
material hardly changes $R$.  Beyond $M_1$ the crushing is so
great that additional mass only serves to make $R$ decrease like
$M^{-1/3}$.  This is the normal white dwarf regime which continues
until the electrons are so confined that they become relativistic, the
radius then shrinks precipitously as the Chandrasekhar limit is
approached.  

The same behaviour is predicted whether we use the crude homogeneous
approximation of Stoner (1930) via (16) or the inhomogeneous ones via
(19) but the detailed numbers will be different.  The fact that for
$M\ll M_1$ the bodies grow at constant density shows that they are not
centrally condensed.  Thus in that regime we should use the
homogeneous model.  However, for the non-relativistic white dwarfs we
should use the polytrope of index ${\scriptstyle{3\over 2}}$, $\Gamma
= {\scriptstyle{5\over 3}}$ and for the relativistic white dwarfs the
polytrope of index 3, $\Gamma = {\scriptstyle{4\over 3}}$.  Each of
these three cases is described by a different value of $\alpha$ which
is ${3 \over 5-n}$ with $n=0$ corresponding to the homogeneous case.
As $n$ varies systematically with mass taking the values $0,\
{\scriptstyle{3\over 2}}\ {\rm and }\ 3$ in the low mass, white dwarf,
and relativistic cases it is not difficult to devise and interpolation
formula for $\alpha$.  However there is one further problem of this
type.  We chose $\zeta$ and $A$ to fit the white dwarf density
profiles but we now find that homogeneous profiles are appropriate for
$M\ll M_1$.  Thus we should interpolate not just $\alpha$ but rather
$(\zeta^2A/\alpha )$ which should take the value $\left
[\zeta^2_oA_o/{\scriptstyle ({3\over 5})}\right ]$ for small $M$ and
$\zeta^2A/\alpha$ with $\alpha = {3 \over 5-n}$ for the white dwarf
regime.  In the whole range the average density increases with mass so
${\zeta^2A \over \alpha G}/(M^{2/3}+M^{2/3}_1)$ decreases as $M$
increases (see 29).

The interpolation
$${\zeta^2A \over \alpha G} = {\zeta^2_oA_o \over 3/5 G} \ \
{M^{2/3}_1 + M^{2/3} \over \left (M^{2/3}_1 + 0.816 M^{2/3} \right )K}\
\ , \eqno (30)$$ where $K=1+{\scriptstyle {3 \over 4}}(1-J^{1/2})$ and
$J=(1-y^4y^{-4}_{ch})^{1/2}$, gives the correct values for $M$ small
and $M$ large and ensures that the density increases.  We now define
$M_p = \left [M^{2/3}_1/0.816 \right ]^{3/2} = 1.356M_1$.  Then
writing $y=(M/M_p)^{1/3}$, our interpolation formula (31) becomes
$${\zeta^2A \over \alpha G} = {5\zeta^2_o A_o \over 3 G} \ \
{M^{2/3}_1 + M^{2/3}\over M^{2/3}_1 \left (1 + y^2  \right ) K}\ \ , \eqno
(31)$$ 
and our Mass radius relationship is
$$\left({\scriptstyle {4 \over 3}}\pi \rho_o \right )^{1/3}R = M^{1/3}_p {y \over 1+y^2 } I \ \ , \eqno (32)$$
with $y=\left ({M \over M_p}\right )^{1/3}$ and $y_{ch}=\left ({M_{ch}
\over M_p} \right )^{1/3}$ and $I = J/K$,
$$\left ({\scriptstyle {4 \over 3}}\pi \rho_o\right )^{1/3} = \beta
{\scriptstyle {5 \over 3}} \left ({\scriptstyle{4 \over 9\pi}}\right
)^{2/3} \mu^{1/3}_e \left ({m_H \over a^3_o}\right )^{1/3} \ \ ; \ \
a_o = {\hbar^2 \over m_ee^2}$$ so $\rho_o$ is apart from structural
constants the density of hydrogen within the Bohr radius, $a_o$, of
the nucleus.  $\rho_o$ depends on $\hbar$ through $a_o$.
$$M_p = 1.356 \ M_1 = 1.356 \left ({\scriptstyle {5 \over 3}}\beta
\right )^{3/2}\mu^{-2}_e \left ({e^2 \over Gm^2_H}\right )^{3/2} m_H \
\ . \eqno (33)$$ 
This is the mass of the cold planet of maximum
radius.  As stated it is essentially the fine structure constant to
the ${\scriptstyle{3 \over 2}}$ power times the Chandrasekhar mass or
alternatively the ratio of electrical to gravitational forces between
two protons, raised to the three halves power, times the hydrogen atom's
mass.

\section{The Mass Radius Relationship}
Our mass-radius relationship is now defined except for the constant
$\beta$ that we inserted in (21)) to allow for the extreme crudeness
of our estimate of the electrical potential energy.  This we shall
evaluate by fitting our formula to the radius of Saturn.  Away from
the relativistic regime formula (32) takes the even simpler form
$$\left ({\scriptstyle{4\over 3}}\pi\rho_o\right )^{1/3}R=M^{1/3}_p \ {y
\over 1+y^2} = {M^{1/3} \over 1+y^2} \ \ . \eqno (34)$$ Now both
$\rho^{1/3}_o$ and $y^{-2}$ are proportional to $\beta$ so for a
planet of known $M$ and $R$ we solve for $\beta $ using Saturn
$$\beta = \left ({\rho \over \rho_o\beta^{-3}} \right )^{1/3} - \left
(\beta y^2 \right )=1.343 - \cdot 230 = 1.113 \ \ , \eqno (35)$$ where
in spite of appearances each bracketed term is independent of $\beta$
and $\rho = M/\left ({4 \over 3} \pi R^3 \right ).$ The same
calculation for Jupiter gives $\beta = 1.161 $ so $\beta = 1.137$ is a
good compromise differing from each by only $2.1$ per cent.  This value of
$\beta$ yields $\rho_o=0.419gm\ cm^{-3}$ and $M_p=6.24 \times
10^{30}gm$.  The density of Hydrogen at the relatively low pressure
(compared with planetary interiors) of half a Megabar is close to 0.54
$gm\ cm^{-3}$ (Alavi et al., 1995).  At much lower pressures solid
Hydrogen and liquid Helium have densities of 0.07 and 0.12 $gm\
cm^{-3}$.  For Terrestial planets $\beta$ (and hence $\rho_o^{1/3}$ and
$M_p^{2/3}$) takes higher values appropriate to 
their composition.  From (34) we see that the planet of maximum radius
has a density of $8\rho_o$.  Thus although $M_p$ is independent of
$\hbar$ nevertheless $R_{\rm{max}}$ depends on $\hbar$.

To summarise our basic result is that the radius $R$ of a cold body of
mass $M$ is given by
$$\left ({\scriptstyle{4 \over 3}}\pi \rho_o \right )^{1/3}R=M^{1/3}_p
{y \over 1+y^2}\ \ \ { \sqrt{1-y^4y_{ch}^{-4}} \over \left \{1+ {3 \over 4} \left
[1-\left (\sqrt{1-y^4y^{-4}_{ch}}\right )^{1/2}\right ]\right \}} \ \ ,
\eqno (36)$$ 
where the surd and the curly bracketed expression reduce to 1 for
non-relativistic white dwarfs and small bodies.

\begin{figure}
\epsfig{file=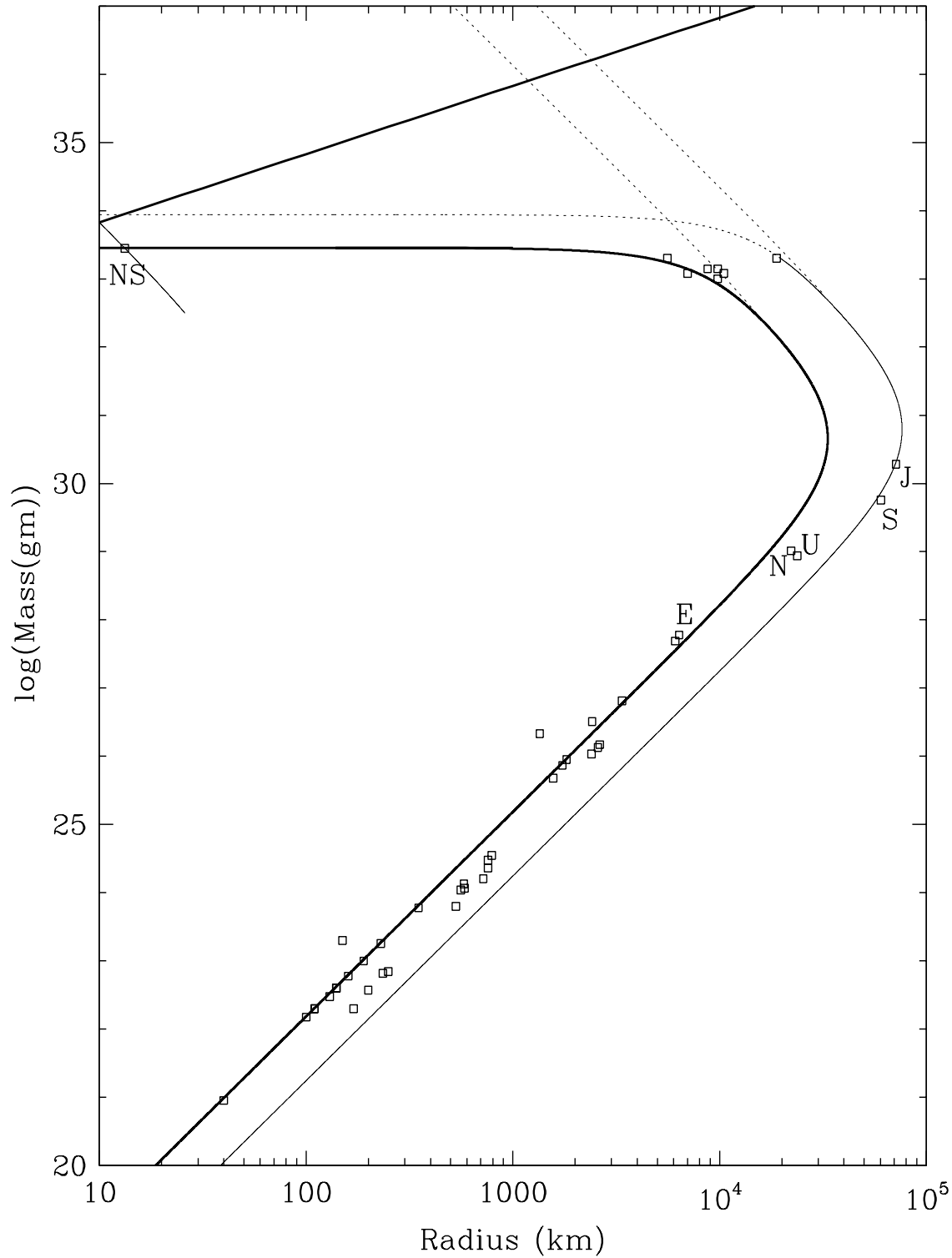,width=15truecm,height=20truecm}
\caption {The dotted line gives the
non-relativistic approximation that omits the surd and the bracketed
expression. The full line gives the full expression for $\rho_o = 3.65
gm cm^{-3},\ \ \mu_e = 2$ appropriate for rocky planets and white
dwarfs.  The lighter line gives the expression for a 25\% Helium 75\%
Hydrogen body or a pristine white dwarf.  The squares are solar system
bodies supplemented by seven white dwarfs of known radius and one
neutron star.  The neutron star line is parallel to the dotted lines
for non-relativistic white dwarfs but reduced in radius from the pure
hydrogen white dwarf by the mass ratio of the neutron and electron.}
\end{figure}

According to (36) the cold Planet of maximum radius has a mass of 3.3
times Jupiter's and its radius is $8.65 \times 10^9 {\rm cms}$ as
opposed to Jupiter's radius of $7.0 \times 10^{9}$ (after correction
at constant volume for the eccentricity).  Equation (36) gives $7.0
\times 10^9$ for the radius of a body of mass $M_J=1.899 \times
10^{30} {\rm gm}$

\section{Neutron Stars}
In the above we have not aimed to treat the mass radius relationship
for neutron stars but with suitable modification the same principles
apply to them.

\begin{enumerate}
\item[I] It is the degenerate neutrons that provide the support so $m_e$ must
be replaced by $m_n$ wherever it occurs.  The electric term is small
and irrelevant.  The mass per neutron is now $m_n$ so $\mu_e$ is
replaced by 1.  In particular under (14) $p_1/\rho_1 = {5 \over
16}c^2$ and in (8) $\rho_1$ is bigger by the factor $\left ({m_n
\over m_e}\right )^3$.
\item[II] The Chandrasekhar mass is almost unchanged except that we are now
interested in $\mu_e=1$ rather than the normal 2.
\item[III]
General Relativistic corrections are no longer so small, so the
effective gravity is somewhat stronger.
\item[IV]
The forces between neutrons are such that the free neutron gas is no
longer a very good approximation.
\end{enumerate}

If we were to ignore all the refinements in III and IV the same
formulae hold provided that $\rho_o$ is replaced by $\left ({m_n \over
m_e}\right )^3\rho_o$ whenever we consider neutron stars.  Thus the
radii are smaller, by the factor ${m_e \over m_n}$, than the white
dwarf of the same mass with $\mu_e=1$.  The Chandrasekhar limit is
almost unchanged at $5.82 M_\odot$. \footnote{Using the values of the
Fundamental Constants \& the Sun's Mass then available Chandrasekhar
(1938) got $5.75 M_\odot$ } In practice effects III and IV reduce this
to between two and 3.6 solar masses, see Shapiro \& Teukolsky.  These
basic points were understood by Baade \& Zwicky (1934) writing soon
after the neutron was discovered. Later they took spectra of the Crab
Pulsar without realising that it pulsed thirty times a second, but
Baade could not interpret the continuous spectrum as, in spite of
Schott's (1912) work, Synchotron radiation was not known in astronomy
at that time.

\appendix
\section[]{Ritter's formula and the evaluation of the $C_n$}
$${1 \over \xi^2}\ \ {d \over d\xi} \left (\xi^2 {d\theta \over
d \xi}\right ) = - \theta^n\ \ , $$
$\theta = 1, \ \ {d \theta \over d \xi} = 0 \ {\rm at}\ \xi = 0$.
Let the edge be at $\xi = \xi_1$.

Define the ``mass'' within $\xi$ by
$\mu(\xi)=\int^\xi_0\theta^n\xi^2d\xi$.

Then $-\xi^2{d\theta \over d \xi} = \mu$ and ${d\mu \over
d\xi} = \xi^2 \theta^n$,

also $\xi^3{d\theta^{n+1} \over
d\xi}=(n+1)\xi^2\theta^n\xi^2{d\theta \over d\xi}/\xi = -
(n+1){\mu \over \xi} \ \ {d\mu \over d\xi}$.

Thus
$$\int^{\xi_1}_0\xi^3 {d\theta^{n+1}\over d\xi} d\xi = -3
\int^{\xi_1}_0 \theta^{n+1}\xi^2d\xi
=-(n+1)\int^{\mu_1}_0{\mu\over \xi}d\mu \ \ , \eqno (A1)$$
Now write $\psi = \theta + \psi_1$ and take $\psi$ to obey
$${1\over \xi^2}\ \ {d\over d \xi} \left (\xi^2 {d \psi \over
d\xi}\right ) = \
	\left \{ 
		\begin{array}{cl}
			- \theta^n  	& \xi \leq \xi_1 \\
 	 		0 		& \xi > \xi_1 
		\end{array}
	\right.
\ \ .
$$
Furthermore choose $\psi_1$ so that $\psi \rightarrow 0$ at $\infty$.
$${d\psi \over d \xi} = \ 
	\left \{ 
		\begin{array}{ll}
			- {\mu \over \xi^2} 	& \xi \leq \xi_1 \\
						& \\
			- \mu_1/\xi^2		& \xi > \xi_1
		\end{array}
	\right.
\ \ .
$$
Hence 
$$\psi_1 = \mu_1/\xi_1 \ \ \ \ \ \psi = \int^\infty_\xi {\mu \over
\xi^2} d\xi \ \ , \eqno (A2)$$
where 
$$\mu = \mu_1 \ {\rm for}\ \xi \geq \xi_1 \ \ ,$$
$$
\begin{array}{rl}
\int^{\xi_1}_0 \xi^2\theta^{n+1}d\xi & = \int^{\xi_1}_0 \theta {d\mu \over
d\xi} d\xi = \int^{\xi_1}_0 (\psi - \psi_1) {d\mu \over d \xi}d\xi =
\int^{\xi_1}_0 \psi {d\mu \over d\xi}d\xi - {\mu_1^2 \over \xi_1}\\
\parbox{20mm}{{\rm using \ }(2)}& \\
& = \int^{\xi_1}_0 \int^\infty_\xi \mu \xi^{-2}d\xi {d\mu \over d \xi} d\xi -{\mu^2_1 \over \xi}= \int^{\mu_1}_0 {\mu \over \xi} d\mu + \int^{\mu_1}_0 {\mu \over \xi} d\mu -{\mu^2_1 \over \xi_1}\\
& \\
& = 2\int^{\mu_1}_0 {\mu \over \xi}d\mu - {\mu^2_1 \over \xi_1} = {n+1 \over 3} \int {\mu d \mu \over \xi}\ {\rm by \ }(1)
\end{array}
$$
Hence 
$${5-n \over 3}\int{\mu d\mu \over \xi}={\mu^2_1 \over \xi_1}$$
which gives Ritter's formula 
$$-{\mathrm V}={3 \over 5-n}\ \ {GM^2 \over R} \ {\rm and }\
\int^{\xi_1}_0 \xi^2 \theta^{n+1} d\xi = {n+1 \over 5-n} \ \ {\mu^2_1
\over \xi_1}\ \ .$$ Now
\medskip
$$
\begin{array}{rl}
C_n & = 		{\int^1_0 3 
	\left (		{\xi \over \xi_1}	\right )
		^2\theta^{n+1}d
	\left (		{\xi \over \xi_1}	\right ) 
	\over 
	\left [		\int^1_0 3 
	\left (		{\xi \over \xi_1}	\right )
		^2 \theta^n d 
	\left (		{\xi \over \xi_1}	\right )
						\right ]
		^{1+{1\over n}}} = 
			{3^{-{1 \over n}} 
			{n+1 \over 5-n} \ \ 
			{\mu^2_1 \over \xi^4_1} \over 
	\left (	\mu_1/\xi^3_1 			\right )
			^{1+{1\over n}}}\\
& \\
C_n & = 3^{-1/n}\ \  {n+1 \over 5-n} \mu_1^{1 - {1 \over n}}\xi^{{3 \over n}-1} \\
& \\
\mu_1 & = -\left (\xi^2 \ \ {d\theta \over d \xi} \right )_{\xi_1}
\end{array}
$$

From tables of polytropes

\begin{center}
\begin{tabular}{lll}
$n = 3/2$ 	& $\xi_1 = 3.65375$ 	& $\mu_1 = 2.71406$ \\
& & \\
$n = 3$		& $\xi_1 = 6.89685$	& $\mu_1 = 2.01824$ \\
& & 
\end{tabular}
\end{center}

Hence
$$
\begin{array}{ll}
C_{3/2} 	& 	= 1.7501 \\
& \\
C_3 		& 	= 2.2146  \\

& 
\end{array}
$$

\label{lastpage}

\end{document}